# Low temperature ferromagnetism, surface paramagnetism and exchange bias effect in nano-grained $\alpha$-Fe$_{1.4}$Ga$_{0.6}$O$_3$ oxide

R.N. Bhowmik[*,1], N. Naresh[1], B. Ghosh[2], and S. Banerjee[2]

[1]Department of Physics, Pondicherry University, R. Venkataraman Nagar, Kalapet, Pondicherry-605014, India

[2]Surface Physics Division, Saha Institute of Nuclear Physics, 1/AF Bidhannagar, Kolkata-64, India

[*]Corresponding author (RNB): Tel.: +91-9944064547; Fax: +91-413-2655734

E-mail: rnbhowmik.phy@pondiuni.edu.in

Abstract

The compound $\alpha$-Fe$_{1.4}$Ga$_{0.6}$O$_3$ has been prepared by mechanical alloying of $\alpha$-Fe$_2$O$_3$ and Ga$_2$O$_3$ and subsequent heating of the mechanical alloyed samples in vacuum condition. Magnetic ordering of $\alpha$-Fe$_2$O$_3$ has been modified by substitution of non-magnetic Ga atoms. The ferromagnetism in $\alpha$-Fe$_{1.4}$Ga$_{0.6}$O$_3$ is more soft and enhanced in comparison to $\alpha$-Fe$_2$O$_3$. Mössbauer spectroscopy also confirmed enhanced ferromagnetic order in $\alpha$-Fe$_{1.4}$Ga$_{0.6}$O$_3$. The room temperature soft ferromagnetism by Ga substitution is interesting for designing $\alpha$-Fe$_2$O$_3$ based unconventional ferromagnet, which is a great challenge to the materials scientists. The $\alpha$-Fe$_{1.4}$Ga$_{0.6}$O$_3$ samples also exhibited features of exchange bias, low temperature surface paramagnetism, and suppression of Morin transition. These features have been affected up to certain extent by the nano-sized grains of the samples.

# 1. Introduction

Ferromagnetic semiconductors have received huge attention in basic sciences for their applications in magnetic recording, magnetic switching, and biomedicines [1, 2]. Materials scientists attempted to develop dilute ferromagnetic semiconductors (DFMS) by substituting small quantity of ferromagnetic atoms, e.g., Fe, Co, Ni, into the lattices of conventional direct band gap semiconductors, e.g., ZnO and GaN [3, 4]. Unfortunately, DFMS do not belong to the class of good ferromagnet [5]. Alternate effort has started to develop non-conventional ferromagnetic semiconductors by suitable substitution of direct band gap semiconductors (e.g., $TiO_2$, $Ga_2O_3$, $Al_2O_3$) into the lattices of antiferromagnetic α-$Fe_2O_3$ [6-11]. The mechanism of non-conventional ferromagnetic semiconductors lies on the altered electro-magnetic properties of rhombohedral planes in α-$Fe_2O_3$ structure as an effect of metal (Ti, Ga, Al, In) substitution. The altered properties in metal substituted α-$Fe_2O_3$ are interesting for the basic sciences, and also promised potential applications in micro-electronics, spintronics, and multi-functional devices [12-14].

After the prediction of a rich magnetic phase diagram in α-$Fe_{2-x}Ga_xO_3$ series by Levine et al. [15], several attempts were made to find the properties of ferromagnetism, band gap tailoring, photoconductivity, and domain switching in α-$Fe_{2-x}Ga_xO_3$ based ferromagnetic semiconductors [16-22]. The current research interest for $Fe_{2-x}Ga_xO_3$ series has been stimulated by prediction of multiferroic properties and tunable ferromagnetic Curie temperature ($T_C$) close to room temperature. Reports [23-27] are mostly available on $FeGaO_3$, a specific composition of $Fe_{2-x}Ga_xO_3$ series, which stabilizes into distorted perovskite structure (orthorhombic). Unfortunately, $FeGaO_3$ is non-suitable for room temperature applications because of its low $T_C$ (~ 225 K). Moreover, $Fe_{2-x}Ga_xO_3$ samples prefer orthorhombic structure at ambient condition [17, 23, 28]. The difficulty of phase stabilization in rhombohedral structure might be the reason for less number of reports on α phase of $Fe_{2-x}Ga_xO_3$ series. In this context, our work [29] on the stabilization of α-$Fe_{1.4}Ga_{0.6}O_3$ in the rhombohedral (α-$Fe_2O_3$) phase is a first step success for developing non-conventional ferromagnetic semiconductors. The technique of mechanical alloying and subsequent non-ambient (vacuum) annealing were found to be effective to form the single-phased rhombohedral structure. Similar technique was later used by R. Saha et al. [30] to stabilize $GaFeO_3$ in rhombohedral phase. The preliminary work on α-$Fe_{1.4}Ga_{0.6}O_3$ [29] showed room temperature ferromagnetism and semiconductor property with optical band gap in the range 2.4-2.5 eV. We believe that room temperature ferromagnetism and optical band gap could be further modified by different Ga substitution in α-$Fe_{2-x}Ga_xO_3$ series. These aspects will be studied in future work. Technologically, such ferromagnetic system can be used for developing advanced spintronics materials, which is suitable for room temperature applications. In this work, we report a detailed study of the low temperature magnetic properties of α-$Fe_{1.4}Ga_{0.6}O_3$ samples. Details of the structural characterization can be found in earlier report [29].

**2. Experimental**

**A. Sample preparation**

The composition α-$Fe_{1.4}Ga_{0.6}O_3$ has been synthesized by mechanical alloying of the mixed powders of high purity α-$Ga_2O_3$ and $Fe_2O_3$ in air using FRITSCH (Pulverisette 6, Germany) planetary mono miller with Tungsten Carbide (5 mm) and steel (10 mm) balls inside the stainless steel bowl. The alloyed powder was taken out after 20 h, 60 h, 80 h, and 100 h milling time. Pellet form of the milled powder was heated at different temperatures under non-ambient (vacuum ~$10^{-6}$ mbar). The single phased compound with α-$Fe_2O_3$ structure (rhombohedral phase with space group R3C) has been stabilized after vacuum heating of the pellet of milled samples at 800 $^0$C for 1 h 40 minutes. The milled samples after vacuum heating at 800 $^0$C were denoted as MA20V8, MA60V8, MA80V8 and MA100V8, respectively (V8 means vacuum heating at 800 $^0$C). In order to check the milling effect on single phased sample, MA100V8 sample (a case study) was made into powder and mechanically milled up to additional 100 h, totaling milling time 200 h including preheated 100 h. The pellet form of these milled samples after vacuum heating was denoted as V8M125, V8M160 and V8M200 for milling time 125 h, 160 h and 200 h, respectively. From analysis of XRD pattern [29], we found grain size (47-15 nm), and cell parameters (a = b: *5.0189*- 5.0343 Å, c: 13.667- 13.740 Å) in rhombohedral structure of the samples. Wavelength dispersive X-ray fluorescence spectrometer (WDXRF: model Bruker S4 pioneer) confirmed atomic ratio of Fe and Ga to 1.4:0.63, which is close to the expected value for $Fe_{1.4}Ga_{0.6}O_3$.

**B. Sample measurements**

SQUID magnetometer (Quantum Design, USA) was used to measure magnetic field (H: 0–70 kOe) and temperature (T: 10 K-310 K) dependent dc magnetization (M). M(T) data were measured using zero field cooled (ZFC) and field cooled (FC) modes. In ZFC mode, the samples were cooled in the absence of external magnetic field (H) from room temperature (300 K) to 10 K, followed by the application of measurement field before starting the record of magnetization on increasing the temperature up to 310 K. After reaching the temperature at 310 K, the sample was field cooled (without changing the measurement field) down to 10 K and magnetization was recorded during the increase of temperature to 310 K. This is denoted as FC mode. M(H) data at selected temperatures were recorded by cooling the samples in ZFC mode. M(H) data for selected samples was also recorded by field cooling the samples to examine the exchange bias effect. $^{57}$Fe Mössbauer spectra were recorded in transmission mode with $^{57}$Co radioactive source in constant acceleration using standard PC-based Mössbauer spectrometer equipped with Wissel velocity drive. Velocity calibration of the spectrometer was done with natural iron absorber at room temperature.

## 3. Results and Discussion

Fig. 1(a) shows the temperature dependence of ZFC magnetization (MZFC(T)) and field cooled magnetization (MFC(T)) of the bulk $\alpha$-Fe$_2$O$_3$ sample at 100 Oe. A typical antiferromagnetic order is noted in $\alpha$-Fe$_2$O$_3$ sample below its Morin transition (T$_M$) ~ 260 K- 270 K without separation between MZFC(T) and MFC(T) curves. The MZFC and MFC separation is seen at T > 270 K, which is canted (anti)ferromagnetic state for $\alpha$-Fe$_2$O$_3$ sample [31]. In contrast, temperature dependence of magnetization curves drastically differ in $\alpha$-Fe$_{1.4}$Ga$_{0.6}$O$_3$ samples [Fig. 1(b-c)]. The separation between MZFC and MFC curves is extended down to 10 K. This indicates an enhancement of ferromagnetic order down to lower temperatures which was previously dominated by anti ferromagnetic order in $\alpha$-Fe$_2$O$_3$. There is a large suppression of the T$_M$ at 270 K in Ga substituted samples and signature of two more Morin transitions has been marked at ~127 K (~160 K for MA20V8) and 43 K. We explain the modified features by considering perturbed antiferromagnetic order and magnetocrystalline anisotropy in Ga substituted $\alpha$-Fe$_2$O$_3$. The magneto crystalline anisotropy (E$_A$) of $\alpha$-Fe$_2$O$_3$ is ~ $K_1^{/}Sin^2\theta$ [32], where $\theta$ is the angle between sublattice magnetization and c (111) axis of rhombohedral plane. The uniaxial anisotropy constant $K_1^{/}$ changes sign from positive below T$_M$ to negative when temperature increases above $T_M$. This is due to competition between two main sources of magnetocrystalline anisotropy in $\alpha$-$Fe_2O_3$ structure, viz., positive signed single ion ($Fe^{3+}$) anisotropy ($K_S$) and negative signed magnetic dipole anisotropy ($K_D$). The fact of $|K_S| > |K_D|$ makes $K^/$ positive below $T_M$ and negative above $T_M$ due to $|K_D| > |K_S|$ [33]. The single ion anisotropy ($K_S$) directs the magnetic easy axis along the c-axis below $T_M$. Hence, spins of the rhombohedral planes lie along out of plane directions, either along c (111) axis for one spins-sublattice or opposite (antiparallel) to c (111) axis for second spins-sublattice. This results in over-all perfect antiferromagnetic spins structure with angle between two magnetic sub-lattices $180^0$ in $\alpha$-$Fe_2O_3$ below $T_M$. On the other hand, dipolar anisotropy ($K_D$) directs the magnetic easy axis within rhombohedral planes. This results in the spins orientation from out of plane direction to in plane direction as the temperature increases above $T_M$. The antiferromagnetic ordering between two spins-sublattices (planes) still is maintained above $T_M$, although spins are ferromagnetically ordered in each sublattice. However, small spin canting between two neighboring planes due to Dzialoshinski-Moriya interactions produces uncompensated ferromagnetization and splitting between MZFC(T) and MFC(T) curves above $T_M$ for $\alpha$-$Fe_2O_3$ sample [34]. We understand that the enhanced ferromagnetism in Ga substituted samples, along with large separation between MFC and MZFC curves down to low temperature is definitely correlated to the decrease of single ion anisotropy ($K_S$) due to replacement of magnetic $Fe^{3+}$ ions by non-magnetic $Ga^{3+}$ ions. The magnetic dipole anisotropy ($K_D$) is expected to be enhanced due to increase of uncompensated ferromagnetic spins between alternate Fe and Ga rich planes. Large separation between FC and ZFC curves below 300 K shows increasing spin frustration in antiferromagnetic planes, which have been perturbed by Ga substitution. The appearance of more than single $T_M$ implies non-uniform perturbation at different rhombohedral planes. The perturbed antiferromagnetic spins order in α-Fe$_2$O$_3$ nanotubes also exhibited two Morin transitions at the temperature range 125 K- 127 K and 237 K-250 K [35]. The observation of better ferromagnetism for the samples with smaller grain size indicated the effect of nano-sized grains on increasing the magnetic dipole anisotropy ($K_D$) in Ga substituted samples [33]. Nano-sized grains affected the magnetic upturn at lower temperatures [Fig. 1(b-c)]. Similar magnetic upturn was noted in Ti substituted α-Fe$_2$O$_3$ samples [36] and understood by surface paramagnetism of antiferromagnetic grains [37]. It may be noted that low field (100 Oe) M(T) curves below 300 K is a collective behavior of different magnetic contributions. A clear competition between antiferromagnetic order arises from cores of the nano-sized grains (showing decrease of MZFC) and ferromagnetic order arises due to uncompensated magnetic order between rhombohedral planes (showing magnetic splitting) and surface paramagnetism arises due to frustrated spins from shell of the grains (showing low temperature upturn).

The nature of M(T) curve has drastically changed at high magnetic field. Insert of Fig. 2 does not show any low temperature magnetic upturn at 70 kOe for α-Fe$_2$O$_3$ sample, confirming typical antiferromagnetic order at low temperatures. The reduction of $T_M$ (~ 230 K) suggests field induced magnetic order in rhombohedral planes. There is no separation between ZFC and FC curves above 230 K. The thermal hysteresis in the temperature range 150 K-230 K, field induced magnetization, and low temperature shifting of $T_M$ suggest first order magnetic transition in α-Fe$_2$O$_3$ during spin reorientation from canted ferromagnetic state (T > $T_M$) to antiferromagnetic state (T < $T_M$) [38]. Fig. 2 shows that the contribution of antiferromagnetic back ground in α-Fe$_{1.4}$Ga$_{0.6}$O$_3$ samples is overcome by high magnetic field (50 kOe) induced magnetization of the planes and magnetic up turn has significantly enhanced due to field induced paramagnetism of the surface spins. A minor signature of thermal hysteresis can be noted for MA20V8 sample below 230 K, but other Ga substituted samples do not show any separation between MFC and MZFC curves (data not shown for clarity) throughout the temperature scale, except they indicated a change of the slope of M(T) curves at 230 K. The change of slope shows a signature of $T_M$ at 230 K, which is prominent in MA20V8 sample and become weak for samples with higher milling time due to better homogeneity of the Ga substitution in $\alpha$-$Fe_2O_3$ structure. The surface paramagnetic contribution brought a significant difference in the nature of the temperature dependence of high field M(T) curves below $T_M$ ~230 K in nano-grained $\alpha$-$Fe_{1.4}Ga_{0.6}O_3$ samples in comparison with the curves of antiferromagnetic $\alpha$-$Fe_2O_3$ sample. The absence of noticeable magnetic splitting indicates that 50 kOe field may be sufficient for saturating ferromagnetic order in Ga substituted samples.

Fig. 3(a) and Fig. 3(b) show M(H) data at 10 K and at 300 K, respectively. The field dependence of magnetization (M(H)) was measured under ZFC mode. The M(H) curve of $\alpha$-$Fe_2O_3$ sample at 10 K shows a typical antiferromagnetic state with up curvature at higher fields and without significant hysteresis loop [33]. On the other hand, M(H) curves with down curvature at higher fields exhibited ferromagnetic character at room temperature (300 K) and at 10 K for $\alpha$-$Fe_{1.4}Ga_{0.6}O_3$ samples. $\alpha$-$Fe_2O_3$ also shows ferromagnetic hysteresis at 300 K. But, ferromagnetization in $\alpha$-$Fe_{1.4}Ga_{0.6}O_3$ samples is large not only at room temperature and also at 10 K. The ferromagnetic hysteresis loop for Ga substituted samples is not clear in the ±70 kOe scale. Low field scale (inset of Fig. 3(a)) shows clear hysteresis loop for Ga substituted samples and the loops are symmetric about the field axis under ZFC mode. This means $\alpha$-Fe$_2$O$_3$ (canted antiferromagnet) has transformed into a good soft ferromagnet at 300 K after Ga substitution in $\alpha$-Fe$_2$O$_3$. The magnetization of $\alpha$-Fe$_{1.4}$Ga$_{0.6}$O$_3$ samples at higher fields has shown increasing trend with milling time (Fig. 3(a-b)), as well as for a particular sample (say, MA100V8 in Fig. 3(c)) with increasing measurement temperature. The lack of magnetic saturation in the M(H) curves of Ga substituted samples at higher fields shows the effect of antiferromagnetic back ground or surface paramagnetism of the nano-sized antiferromagnetic grains [37]. These aspects will be elucidated by the analysis of M(H) curves. The minor irreversibility between M(H up) and M (H down) curves with applied field up to 50-60 kOe shows that the spins may not be completely relaxed in during field induced reorientation process. For $\alpha$-Fe$_2$O$_3$ sample, the up curvature in M(H > 20 kOe) curves with minor irreversibility at 10 K is different from the nearly linear and reversible M(H> 20 kOe) curves at 300 K. As shown in the inset of Fig. 3(b), differential magnetic susceptibility ($\delta M/\delta H$) at 10 K was increasing at field > 20 kOe. This shows a strong field induced spin reorientation process in the antiferromagnetic state of $\alpha$-Fe$_2$O$_3$ with typical spin flop field ~ 20 kOe and comparable to reported value [33, 34]. Such spin reorientation process becomes weak at 300 K where $\alpha$-Fe$_2$O$_3$ exhibited canted ferromagnetic state and differential magnetic susceptibility decreases with field > 20 kOe. The minor irreversibility in the high field M(H) curve is also noted in $\alpha$-Fe$_{1.4}$Ga$_{0.6}$O$_3$ samples, particularly at 10 K. The differential magnetic susceptibility decreases with field (> 20 kOe) for the M(H) curves at 10 K and 300 K (inset of Fig. 3(c)). The decrease of differential magnetic susceptibility with field (>20 kOe) gives another evidence of enhanced ferromagnetism in Ga substituted samples.

The ferromagnetic parameters (coercivity ($H_C$), spontaneous magnetization ($M_S$)) of the samples have been calculated using M(H) loop. $H_C$ is the average (($|H_{C1}|+ |H_{C2}|$)/2) of the fields on negative ($H_{C1}$) and positive (($H_{C2}$) field axis where magnetization is zero (see inset of Fig. 3(a)). The high field side of M(H) curves of Ga substituted samples showed a superposition of paramagnetic or residual antiferromagnetic component (reversible and almost linear) along with irreversible ferromagnetic loop at lower field. The spontaneous magnetization in such case was determined by extrapolating the linear M(H:70 to 0 kOe) curve from higher field side to H = 0 value on M axis, and applying Arrot plot ($M^2$ vs. H/M) on the initial M (H:0 to 70 kOe) curves (see Fig. 4(a) for 10 K and 300 K data). The spontaneous magnetization ($M_S$) calculated using extrapolation of high field M(H) curve is close to the value obtained from polynomial fit of the Arrot plot (inset of Fig. 4(a)). Hence, the value of $M_S$ has been calculated by averaging the values obtained from two methods. The non-linear (up curvature shaped) increase in Arrot plot is related to the field induced magnetic ordering of canted spins [39]. The contribution of surface paramagnetic susceptibility ($\chi_{sp}$) of the nano-sized grains has been calculated by extrapolating the most linear portion of the Arrot plot at higher field side on the H/M axis (see inset of Fig. 4(a)). The $\chi_{sp}$ of Ga substituted samples ( $(7.2-12.0) \times 10^{-5}$ emu/g/Oe) is significantly large in comparison with the value (($1.6 - 2.0) \times 10^{-5}$ emu/g/Oe) reported by Bercoff et al. [34] for $\alpha$-$Fe_2O_3$ nanoparticles, and also in comparison with $\chi_{sp}$ ($\leq 2.2 \times 10^{-5}$ emu/g/Oe at 300 K) in our $\alpha$-$Fe_2O_3$. The $\chi_{sp}$ values for the Ga substituted samples at with milling time have been shown in the inset of Fig. 4(b) at measurement temperatures 10 K, 200 K and 300 K. The general tendency is that $\chi_{sp}$ increases with milling time, except some fluctuated behavior for milling time less than 100 hrs where milling was performed before vacuum annealing. At the same time, the values of $\chi_{sp}$ at 200 K are smaller in comparison to the values at 10 K and 300 K and the feature is independent of milling times. The temperature dependence of $\chi_{sp}$ in MA100V8 sample (Fig. 3(c)) has confirmed rapid increase of surface paramagnetic susceptibility at lower temperatures. The minimum of $\chi_{sp}(T)$ curve at about 100 K indicates a competition between surface paramagnetism, which dominates below 100 K and antiferromagnetic back ground from core of the grains, which comes into play below Morin transition ~ 230 K. Although signature of Morin transition (~ 230 K) was indicated in high field M(T) curves of MA100V8 sample (Fig. 2), but $\chi_{sp}(T)$ curve seems to be more effective for distinguishing the features of surface paramagnetism and antiferromagnetic contribution in the samples.

The values of $M_S$ and $H_C$ of the Ga substituted samples are shown in Fig. 5(a) and Fig. 5(b), respectively. The results suggest enhancement of soft ferromagnetic character (large $M_S$ and low $H_C$) in $\alpha$-Fe$_{1.4}$Ga$_{0.6}$O$_3$ samples in comparison with bulk $\alpha$-Fe$_2$O$_3$ sample with low $M_S$ (~0.292 emu/g) and large $H_C$ (~3910 Oe). In Ga substituted samples, the $M_S$ has increased with milling time up to 60 h and then, decreased on further increase of the milling time. On the other hand, $H_C$ has decreased initially on increasing the milling time up to 60 h and then, significantly increased for milling time 100 h. Thereafter, $H_C$ slowly decreased on further increase of the milling time. The trend of the variation of $M_S$ and $H_C$ is same for all the measurement temperatures 10 K, 200 K, and 300 K. Using the relation $K_1^{/} \sim (H_C \times M_S)/2$, we noted an increasing trend of anisotropy constant ($K_1^{/}$) up to milling time 100 hrs, which could be seen as the affect of mechanical alloying. In the samples where milling was performed after vacuum annealing at 800 $^0$C, small fluctuation was characterized by a slight decrease of $K_1^{/}$ at 160 hrs milling time and again increase at 200 hrs milling time. The trend of the variation of $K_1'$ with milling time before vacuum heating and after heating is irrespective of the measurement temperatures at 10 K, 200 K and 300 K. The $K_1'$ value is smaller at 200 K in comparison with 300 K and 10 K. Interestingly, $K_1'$ value at 10 K is higher than the value at 300 K. This is due to extra anisotropy contribution from surface spins of the nano-sized grains, dominating at lower temperatures. Exact nature of the temperature dependence of $K_1'$ has been understood from the temperature dependence of $M_S$ (inset of Fig. 5(a)) and $H_C$ (inset of Fig. 5(b)) for MA100V8 sample. The $M_S(T)$ has decreased with increasing temperature from 10 K (~1.30 emu/g) to minimum value (~0.97 emu/g) at 150 K, followed by an increase to 1.02 emu/g and 1.07 emu/g at 200 K and 300 K, respectively. Similarly, $H_C(T)$ has decreased with increasing temperature from 10 K (~248 Oe) to minimum (~151 Oe) at 150 K, followed by an increase of $H_C$ with temperature up to 300 K. We found that all ferromagnetic parameters ($M_S$, $H_C$, $K_1'$) attained minimum at ~ 150 K and there is a good connectivity among the temperature dependence of $\chi_{sp}$, $M_S$ and $H_C$. We conclude that surface magnetism dominates below 150 K, where as uncompensated ferromagnetism dominates above 150 K and 150 K is a competitive zone for the major contributors to anisotropy and magnetism. In addition to increasing surface anisotropy, the antiferromagnetic back ground acts as the pinning centers or hard magnetic phase for increasing coercivity below 150 K. We also noted that the surface paramagnetic susceptibility throughout the measurement temperatures 10 K-300 K is connected to the ferromagnetization of nano-grain sized MA100V8 sample. This is consistent to the fact that perturbation of shell spin structure is largely responsible for ferromagnetism and surface paramagnetism in antiferromagnetic grains [37]. Wang et al. [7] reported similar trend of the temperature dependent $H_C$ in $\alpha$-$Fe_{1.2}Ga_{0.8}O_3$. However, our samples are ferromagnetically more soft ($H_C$ < 250 Oe at 10 K and 300 K) than $\alpha$-$Fe_{1.2}Ga_{0.8}O_3$ sample ($H_C$~800 Oe at 5 K). The lower value of $H_C$ in our samples arises due to higher Fe content in comparison with $\alpha$-$Fe_{1.2}Ga_{0.8}O_3$. The increase of magnetic softness was reported in single crystals of $Ga_{2-x}Fe_xO_3$ with increasing Fe content (x) [19]. Mukherjee et al. [17] observed maximum $H_C$ in polycrystalline $\alpha$-$Ga_{2-x}Fe_xO_3$ for x = 1. The increase of in-plane uniaxial anisotropy in our Ga substituted samples is indicated from the elongated ferromagnetic loop along the magnetization axis and simultaneous reduction of loop width along field direction [40, 41].

Fig. 6 (a-e) compared the M(H) loops at 10 K for different Ga substituted samples, which have been measured under ZFC mode and FC mode at 50 kOe cooling field. We see an appreciable shift of the M(H) loop under FC mode with respect to ZFC mode. This suggests exchange bias effect in Ga substituted samples. Exchange bias shift ($\Delta H_{exb} = H_0^{FC} - H_0^{ZFC}$) has been calculated from the shift of the center [$H_0^{FC} = (H_{C1}^{FC} + H_{C2}^{FC})/2$] of the FC loop with respect to the center [$H_0^{ZFC} = (H_{C1}^{ZFC} + H_{C2}^{ZFC})/2$] of the ZFC loop. Fig. 6(f) shows negative exchange bias field for all Ga substituted samples. This shows sufficiently strong antiferromagnetic interactions among the rhombohedral planes of $\alpha$-$Fe_{1.4}Ga_{0.6}O_3$ samples, despite the fact that ferromagnetism has been enhanced after Ga substitution. The exchange bias shift ($\Delta H_{exb}$) is the lowest for the sample with milling time 60 hours. A slow increase of $\Delta H_{exb}$ for the samples with milling ≥ 100 hours is most probably related to better magnetic homogeneity. In $\alpha$-$Fe_{1.4}Ga_{0.6}O_3$, Fe rich and Ga rich layers with different magnitude of magnetic order is expected. Most probably, $Ga^{3+}$ ions are randomly substituted $Fe^{3+}$ ions in rhombohedral planes. In addition to nano-sized grains, change of magnetic anisotropy and non-ambient (vacuum) heating affected the exchange coupling among different magnetic planes in $\alpha$-$Fe_{1.4}Ga_{0.6}O_3$ samples. Some of these planes preferably ordered along the cooling field direction than the others, and exhibited exchange bias shift. The exchange bias effect in metal substituted $\alpha$-$Fe_2O_3$ system was also explained due to exchange coupling between different magnetic layers and variation of magnetic anisotropy [42-45].

    Magnetic ordering of the Ga substituted $\alpha$-$Fe_2O_3$ samples at microscopic level was checked using Mössbauer spectroscopy. Room temperature spectra (Fig. 7) were analyzed with NORMOS-SITE program. The spectra were fitted with (ferromagnetic) sextet and one single (paramagnetic) line. The paramagnetic component appeared due to small grains of the samples. The values of fit parameters are shown in Table 1. The fraction of paramagnetic component has increased in MA200 in comparison with MA60 due to decrease of grain size [29]. Values of Isomer shift are characteristic of $Fe^{3+}$ state in all samples. The fit parameters indicated some changes at the microscopic level of Ga substituted samples. One reason for less variation of IS and QS values is that Ga and Fe have same valence state +3 and ionic radius is nearly same. Hence, less distortion is expected in of $Fe_2O_3$ after non-magnetic Ga substitution. The six-line spectrum of Ga substituted $\alpha$-$Fe_2O_3$ samples is similar to the spectrum of $\alpha$-$Fe_2O_3$ [46]. The identical magnetic structure in Ga substituted $\alpha$-$Fe_2O_3$ and $\alpha$-$Fe_2O_3$ samples are also noted (not shown) in Mössbauer spectra at 5 K under 50 kOe field. However, the observed spectrum is different from the spectrum of ferrimagnetic ordering suggested in $GaFeO_3$ [45].

## 4. Conclusions

Ferromagnetization in $\alpha$-$Fe_{1.4}Ga_{0.6}O_3$ compound has been enhanced in comparison with $\alpha$-$Fe_2O_3$. Ga substitution in $\alpha$-$Fe_2O_3$ structure indicated a significant decrease of single ion ($Fe^{3+}$) anisotropy and increase of magnetic dipole anisotropy. However, single ion anisotropy is strong enough to compete with magnetic dipole anisotropy for retaining the weak signature of $T_M$ at ~270 K (100 Oe) and at ~ 230 K (50-70 kOe). In addition to the Ga substitution effect, low temperature magnetic properties have been affected by nano-sized grains of the samples. Low field M(T) curves showed a competition between antiferromagnetic core (showing decrease of MZFC) and surface paramagnetism due to shell of the nano-sized grains (showing low temperature upturn) and soft ferromagnetism due to perturbed antiferromagnetic order in rhombohedral planes (showing magnetic splitting). The temperature dependence of paramagnetic susceptibility, calculated from Arrot plot of the M(H) data, distinguished the contribution of surface paramagnetism, dominating at low temperature, from the antiferromagnetic back ground of the grains, dominating below Morin transition. Magnetic uncompensation between Fe rich and Ga rich layers has produced enhanced ferromagnetism, whereas exchange coupling between different layers and variation of anisotropy exhibited exchange bias at low temperature. The development of soft ferromagnetism in $\alpha$-$Fe_{1.4}Ga_{0.6}O_3$ could be a major success for realizing the $\alpha$-$Fe_2O_3$ based ferromagnet for room temperature applications.

## Acknowledgment

RNB acknowledges research grant from DST (NO. SR/S2/CMP-0025/2011) and CSIR (No. 03(1222)/12/EMR_II), Govt. of India. Authors also thank to Dr. V.R. Reddy of UGC-DAE CSR, Indore Centre for recording Mössbauer spectra of the samples.

**Figure Captions**

Fig.1 (a-d) Temperature dependence of ZFC and FC magnetization at 100 Oe for $\alpha$-$Fe_2O_3$ and Ga substituted $\alpha$-$Fe_2O_3$ samples. Dotted lines indicate the signature of Morin transitions.

Fig. 2 Temperature dependence of magnetization for different samples measured in ZFC and FC modes at 50 kOe field. Inset shows the M(T) curves (ZFC and FC) measured at 70 kOe for $\alpha$-$Fe_2O_3$. Dotted lines represent the possible Morin transitions.

Fig.3 (Color online) M(H) loop of different samples at 10 K (a) and 300 K (b), at 10 K-300 K for MA100V8 sample (c). Insets show magnified M(H) loop (a) and $\delta M/\delta H$ (b, c)

Fig.4 (Colour online) (a) $M^2$ vs H/M plot using first quadrant of M(H) loop at 300 K and 10 K in ZFC mode for selected samples. Inset shows the calculation of $M_S$ by polynomial fit. Temperature dependent $\chi_{sp}$ for MA100V8 sample is shown in (b). Inset shows the variation of $\chi_{sp}$ with milling time at selected temperatures.

Fig. 5 (a) Milling time dependence of spontaneous magnetization $M_S$ (in a) and Coercivity ($H_C$) (in b) at different temperatures. Insets of (a) and (b) show the temperature variation of $H_C$ and $M_S$, respectively, for MA100V8 sample.

Fig. 6 (Color online) M(H) loop at 10 K measured in ZFC and FC modes for different samples [in (a)-(e)]. Exchange bias field of the samples with milling are shown in (f).

Fig. 7 (Colour online) Room temperature Mössbauer spectra of selected samples. Lines show fit of the spectra.

Table 1. Obtained values of Hyperfine parameters (full width at half maximum (FWHM), Isomer shift (IS), Quadrupole splitting (QS), hyperfine magnetic field (BHF), area ratio of second and third lines in a sextet ($A_{23}$)) from the fit of room temperature Mössbauer spectra.

| Samples | Fitted spectrum component | FWHM (mm/s) | IS (mm/s) | QS (mm/s) | BHF (Tesla) | $A_{23}$ (%) |
|---|---|---|---|---|---|---|
| MA60 | ferromagnetic | 0.549±0.02 | 0.364±0.01 | 0.185±0.012 | 50.75±0.04 | 76.8 |
| | paramagnetic | 1.41±0.11 | 0.34±0.04 | --- | --- | 23.2 |
| MA100 | ferromagnetic | 0.495±0.02 | 0.368±0.01 | 0.186±0.012 | 50.78±0.04 | 71.6 |
| | paramagnetic | 1.55±0.12 | 0.276±0.04 | --- | --- | 28.3 |
| MA160 | ferromagnetic | 0.50±0.02 | 0.366±0.01 | 0.184±0.016 | 50.83±0.06 | 72.1 |
| | paramagnetic | 1.48±0.13 | 0.41±0.04 | --- | --- | 27.9 |
| MA200 | ferromagnetic | 0.53±0.02 | 0.347±0.01 | 0.159±0.013 | 50.75±0.05 | 68.9 |
| | paramagnetic | 1.46±0.09 | 0.32±0.03 | --- | --- | 31.1 |

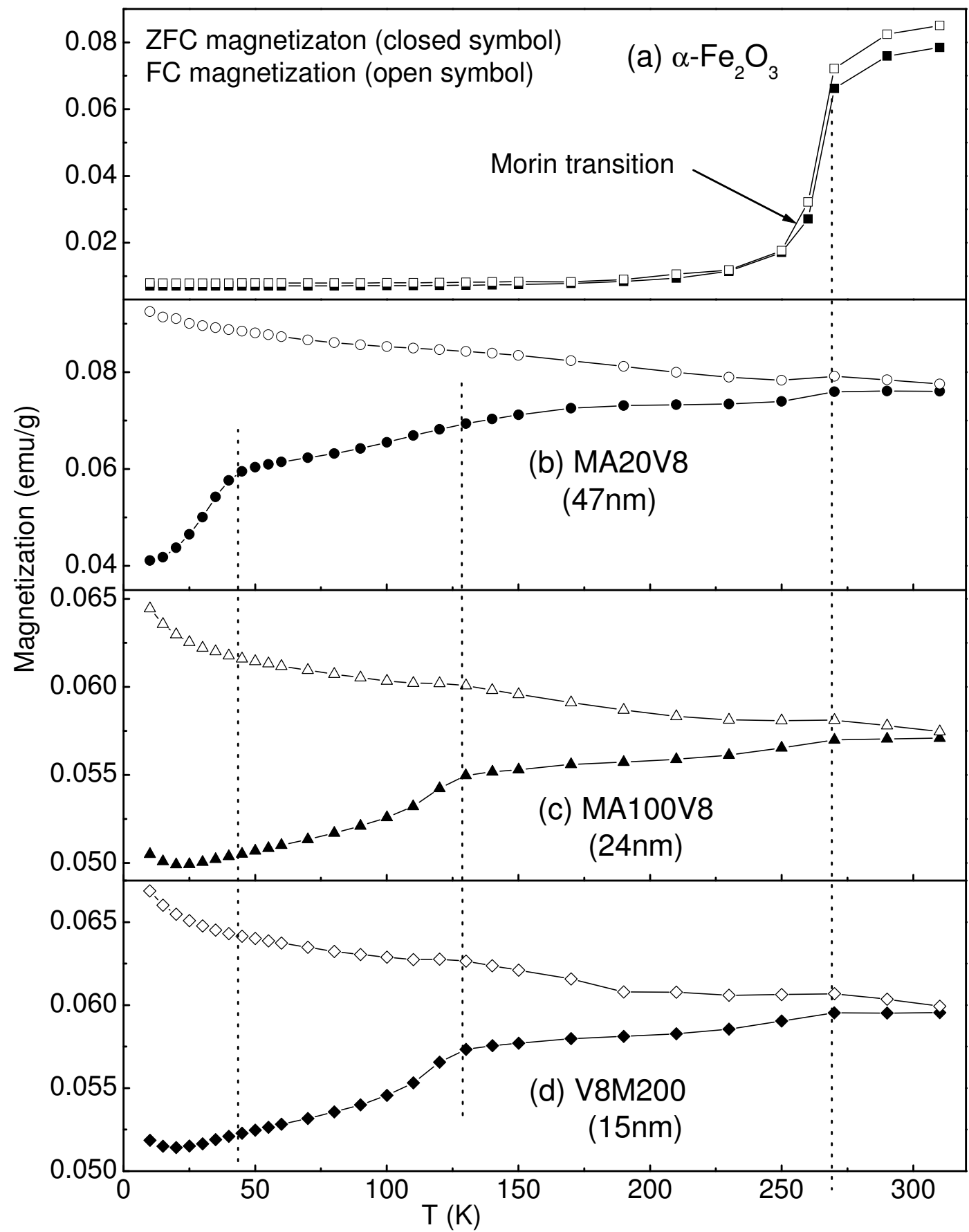

Fig.1 [(a)-(d)] Temperature dependence of ZFC and FC magnetization at 100 Oe for $\alpha$-$Fe_2O_3$ and Ga doped $\alpha$-$Fe_2O_3$ samples. Dotted lines indicate the signature of Morin transitions.

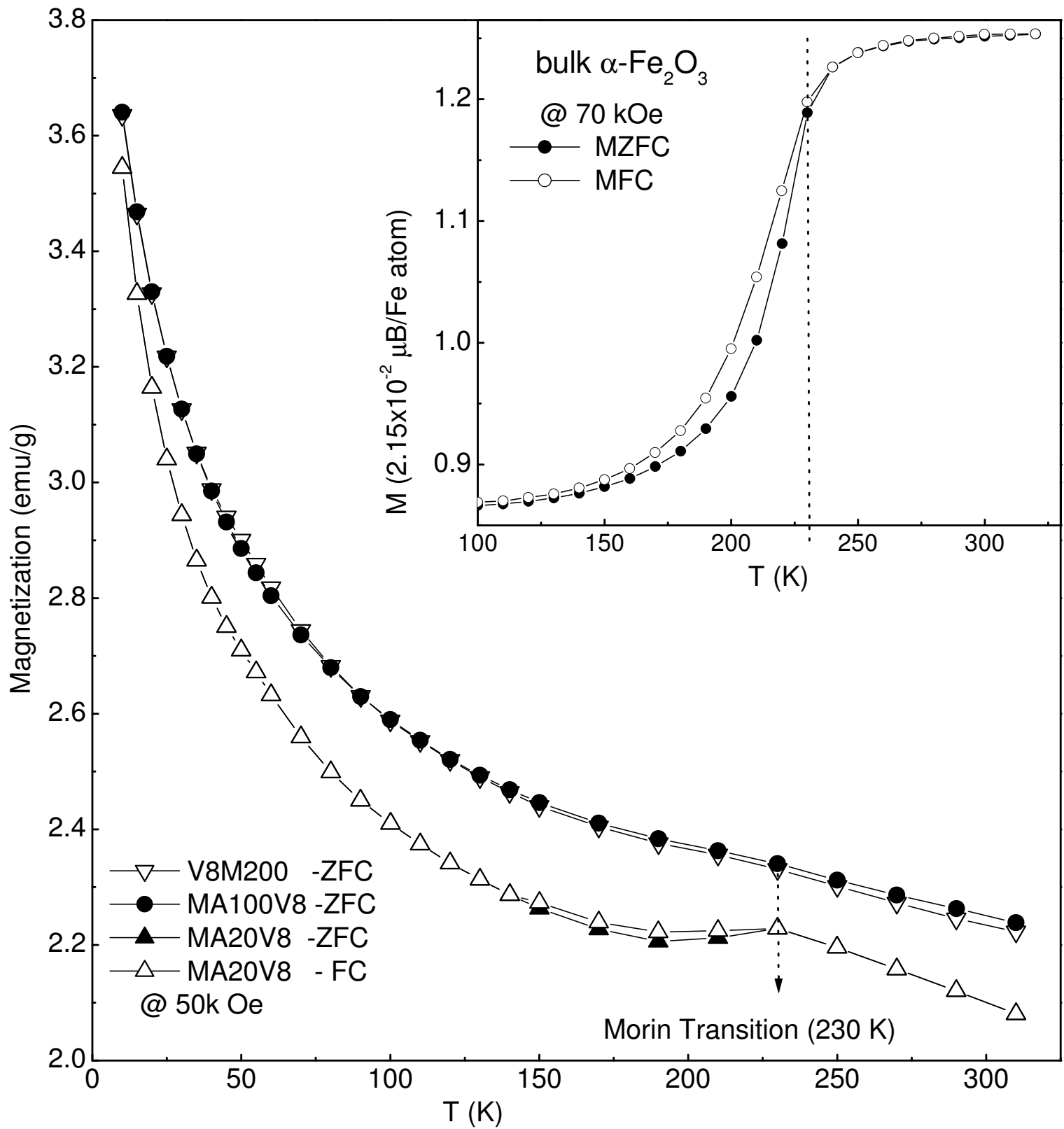

Fig. 2 Temperature dependence of magnetization for different samples measured in ZFC and FC modes at 50 kOe field. Inset shows the M(T) curves (ZFC and FC) measured at 70 kOe for $\alpha$-$Fe_2O_3$. Dotted lines represent the possible morin transitions.

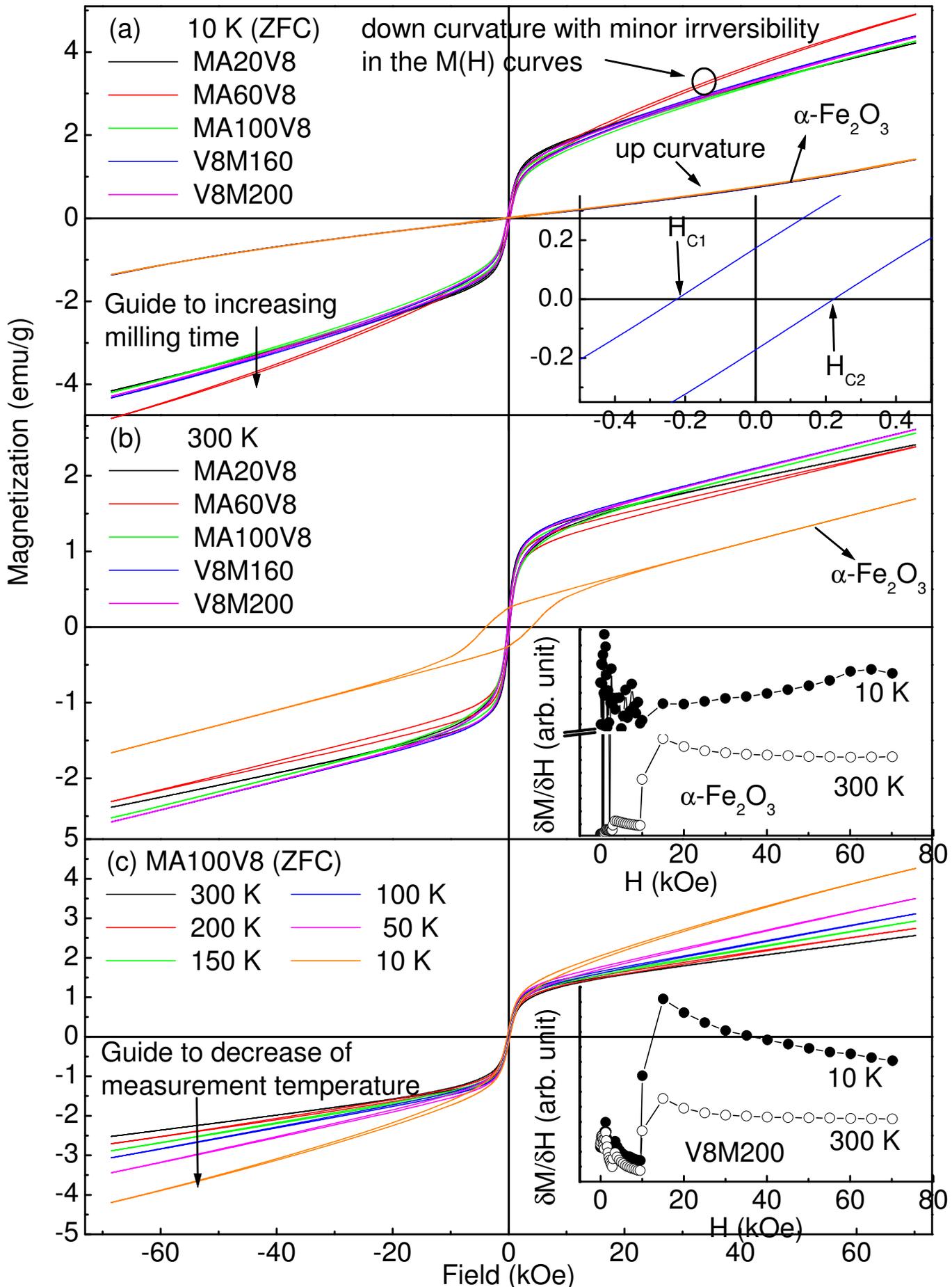

Fig.3 (Color online) M(H) loop of different samples at 10 K (a) and 300 K (b), at 10 K-300 K for MA100V8 sample (c). Insets show magnified M(H) loop (a) and δM/δH (b,c)

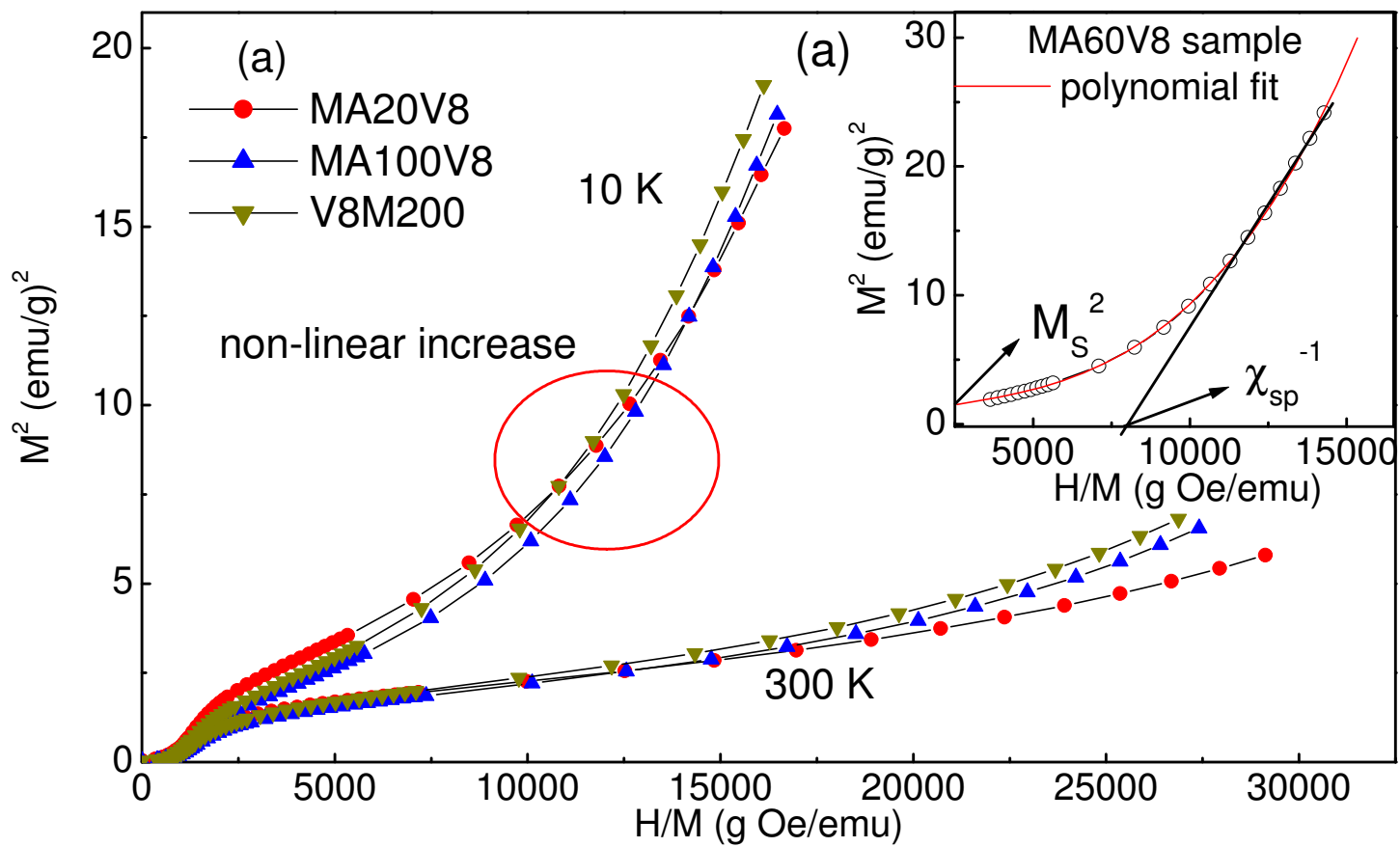

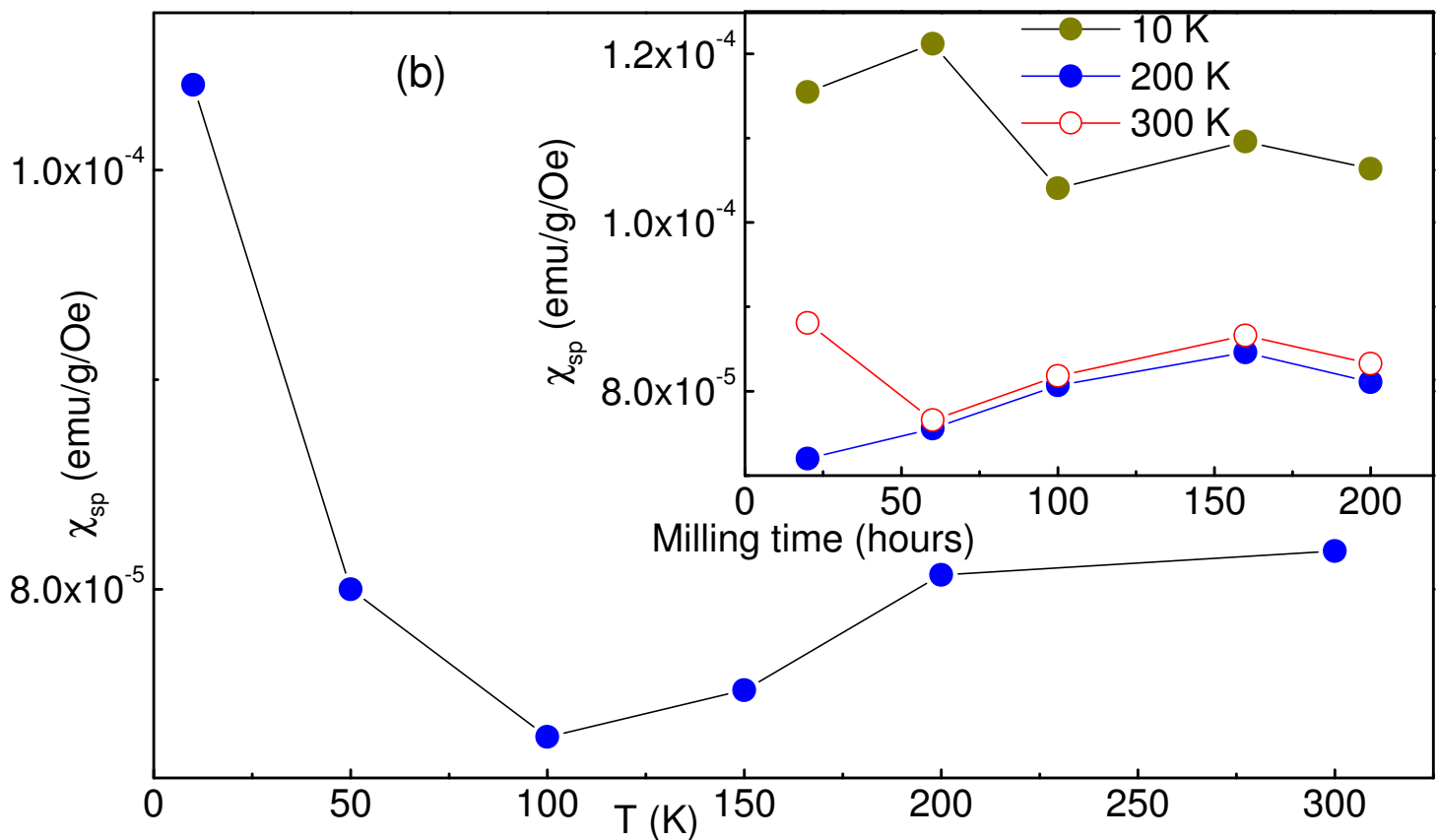

Fig. 4 (Colour online) (a) $M^2$ vs. $H/M$ plot using first quadrant of $M(H)$ loop at 300 K and 10 K in ZFC mode for selected samples. Inset shows the calculation of $M_s$ by polynomial fit. Temperature dependent $\chi_{sp}$ for MA100V8 sample is shown in (b). Inset shows the variation of $\chi_{sp}$ with milling time at selected temperatures.

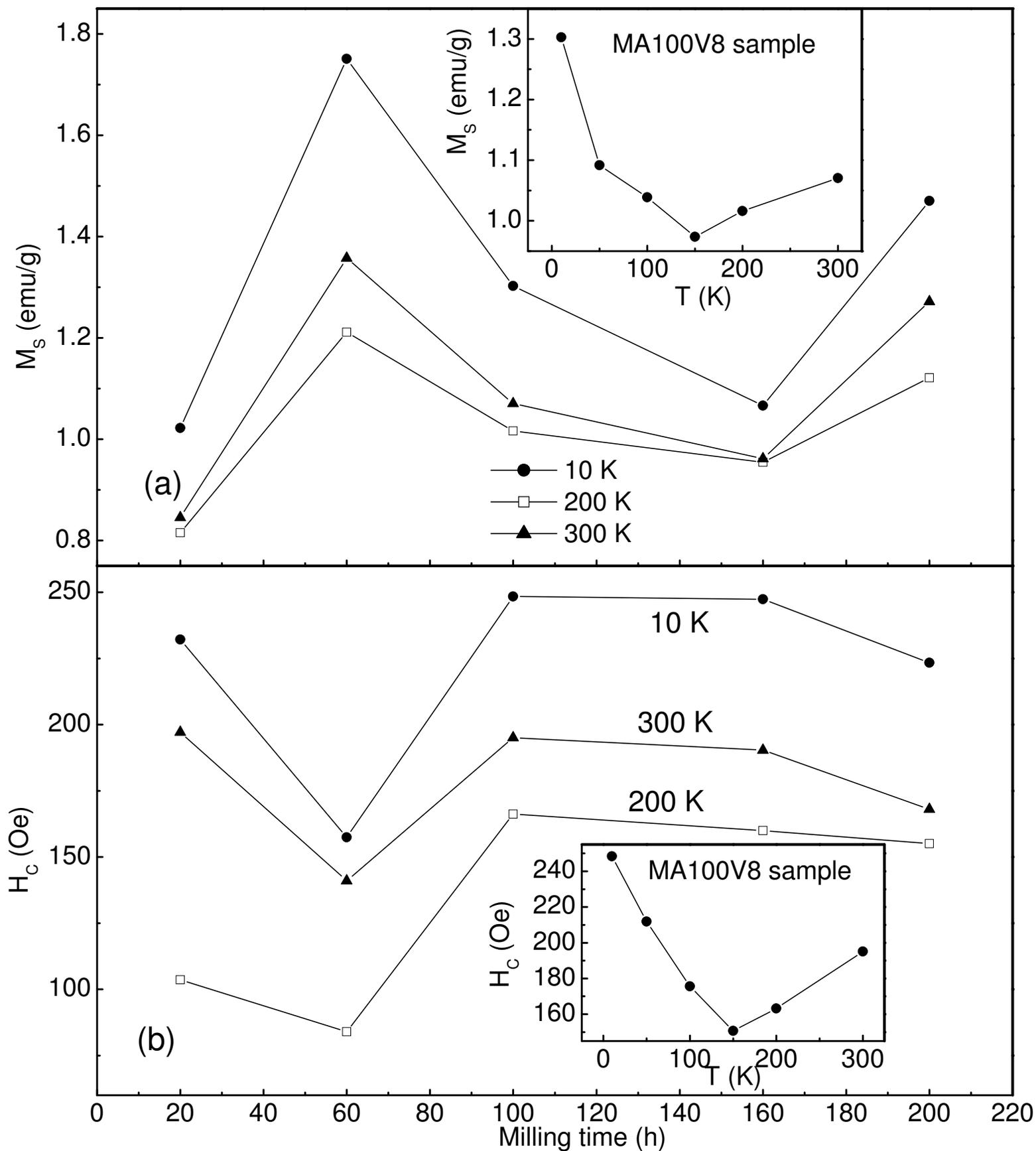

Fig. 5 Milling time dependence of spontaneous magnetization ($M_S$) (in a) and Coercivity ($H_C$) (in b) at different temperatures. Insets (a) and (b) show the temperature variation of $H_C$ and $M_S$, respectively, for MA100V8 sample.

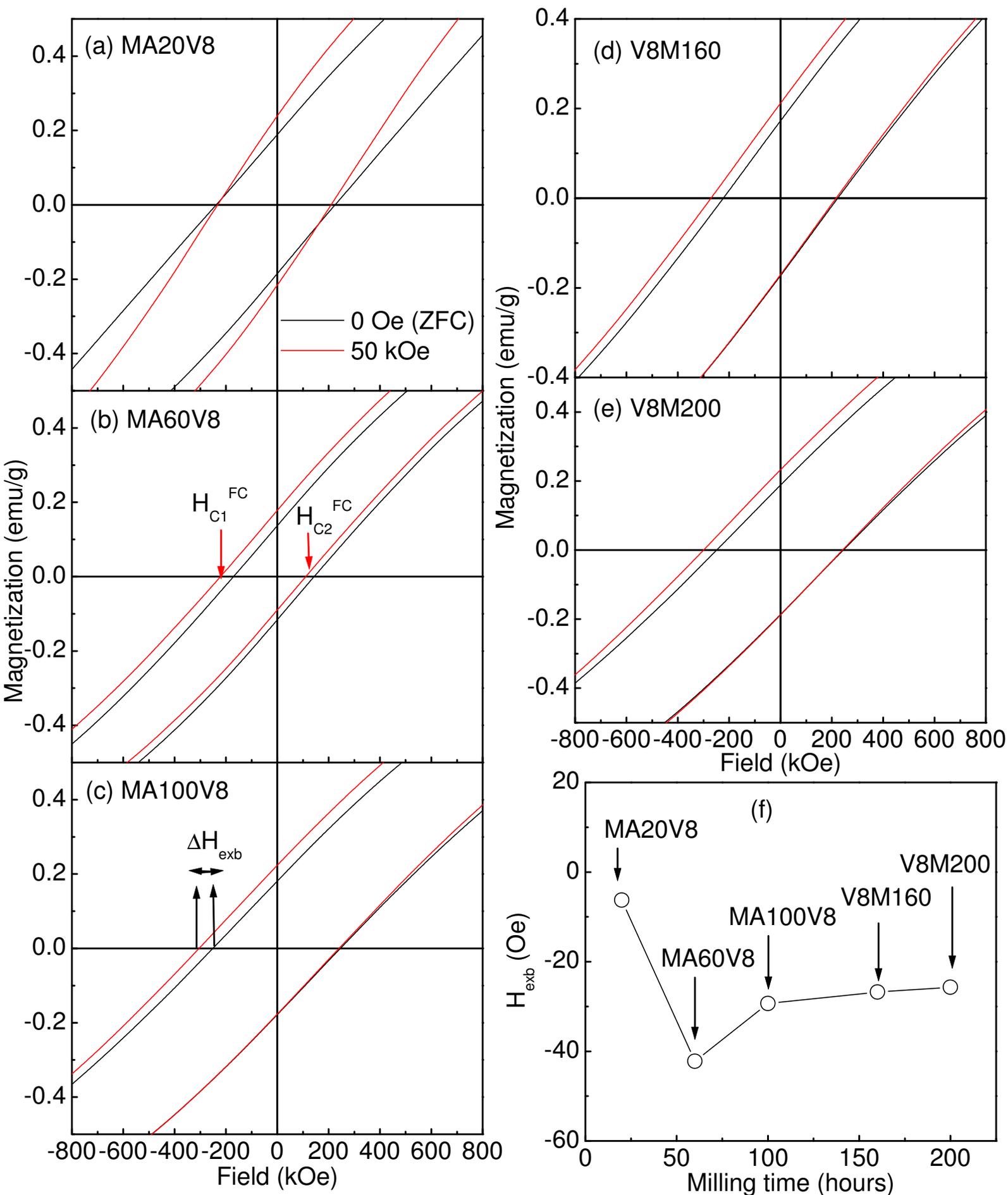

Fig.6 (Color online) M(H) loop at 10 K measured in ZFC and FC mode for different samples [in (a)-(e)]. Exchange bias field of the samples with milling are shown in (f).

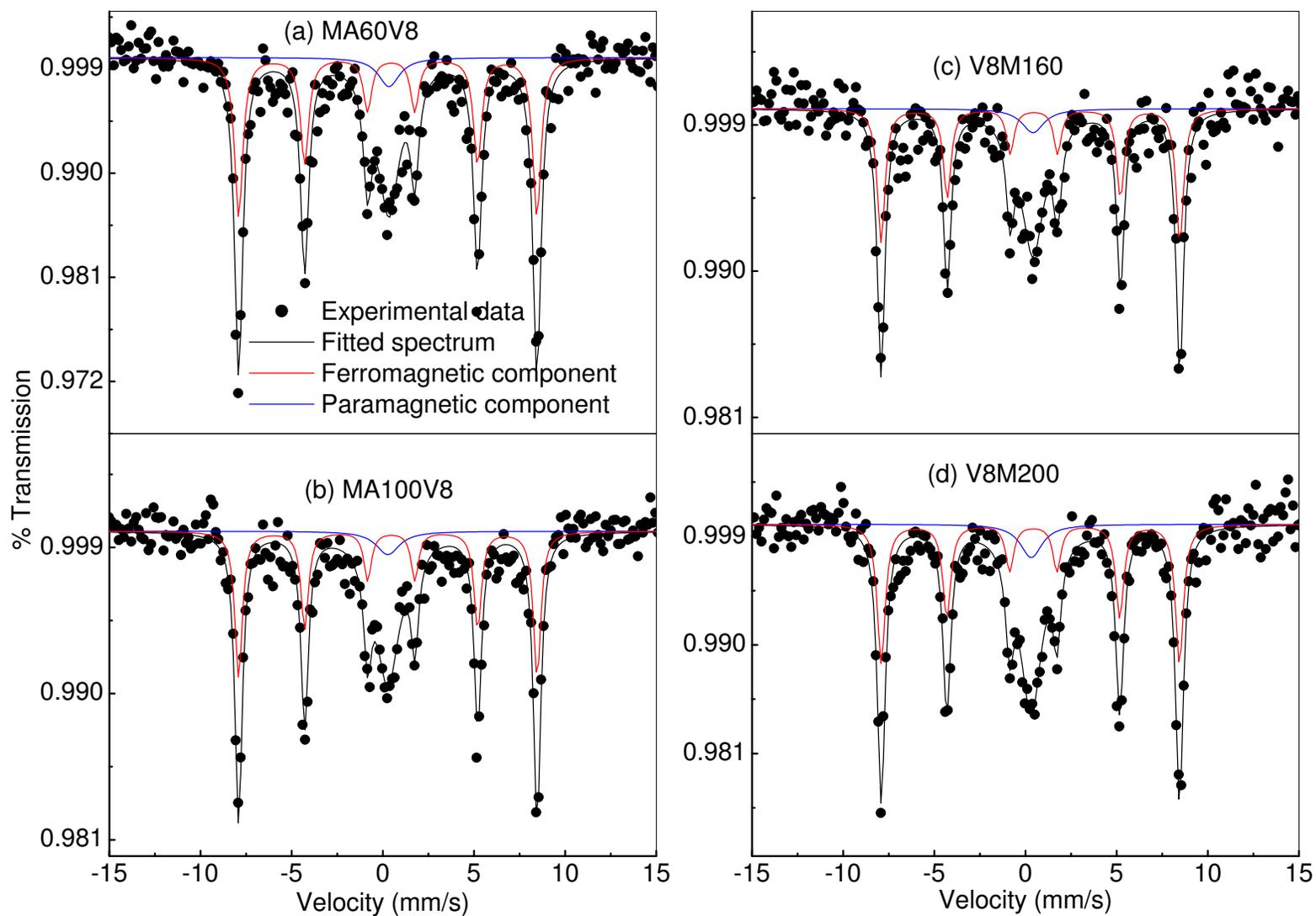

Fig.7 (Colour online) Room temperature Mossbauer spectra of selected samples. Lines show fit of the spectra.